\documentclass[runningheads,a4paper]{llncs}

\usepackage{amssymb}
\usepackage{graphicx}
\usepackage{float}
\usepackage[singlelinecheck=true]{caption}
\usepackage[section]{placeins}
\usepackage{tabularx,ragged2e,booktabs}
\usepackage{ctable}
\usepackage{multirow}
\usepackage{mathtools}

\newcommand{\keywords}[1]{\par\addvspace\baselineskip
\noindent\keywordname\enspace\ignorespaces#1}

\begin{document}

\mainmatter  

\title{Non-trivial reputation effects on social decision making in virtual environment}

\titlerunning{Non-trivial reputation effects on social decision making }
\author{Mirko Duradoni$^3$\and Franco Bagnoli$^1$ $^2$\and Andrea Guazzini$^2$ $^3$ }

\institute{
$^1$Department of Physics and Astronomy, University of Florence, Florence, Italy\\
$^2$Center for the Study of Complex Dynamics, University of Florence, Florence, Italy\\
$^3$Department of Science of Education and Psychology, University of Florence, Florence, Italy }
\maketitle

\begin{abstract} Reputation systems are currently used, often with success, to ensure the functioning of online services as well as of e-commerce sites. Despite the relationship between reputation and material cooperative behaviours is quite supported, less obvious appears the relationship with informative behaviours, which are crucial for the transmission of reputational information and therefore for the maintenance of cooperation among individuals. The purpose of this study was to verify how reputation affects cooperation dynamics in virtual environment, within a social dilemma situation (i.e., where there are incentives to act selfishly). The results confirm that reputation can activate prosocial conducts, however it highlights also the limitations and distortions that reputation can create. 

\end{abstract}

\section{Introduction}
The cooperative behaviours have been and are still the object of study of many disciplines, including evolutionary biology, antropology, sociology, social psychology and sociophysics. 
This interest arises from the fact that this type of conduct apparently seems to fall outside the natural selection theory. In addition, social behaviours appear in contrast, especially when these entail costs for the one who puts them in place, with the view of humans as self-interested agents (\textit {i.e., homo oeconomicus}). Indeed humans appear to regulate their own behaviour in accordance with rules and standards different from rationality \cite{Lewin}. In particular, most of our actions are determined by the social environment, or, better, by what we perceive as our social environment. 
Laboratory \cite {Haley05} and field studies \cite{Bate06} \cite{Ernest11} have shown that subtle cues of the presence of other people, such as a photograph or a synthetic eyes on a computer screen, alter, in an almost unconscious way, our prosocial behaviour, as well as our performance and physiological activation. 
Human beings are therefore deeply influenced by the social environment in which they live. 
This influence according to the \textit{“social heuristics hypothesis”} \cite{Rand14} leads to the development of intuitive and economic models of decision-making (\textit{i.e.,}  heuristics) adaptive in the social context. 
Rand and his colleagues have shown that humans have a generalized tendency to cooperate, which however can be overridden by deliberation. 
It is therefore essential to identify the factors that promote cooperation. One of the mechanisms that seem to be able to maintain cooperation among humans is reputation \cite {Alex87}. Both computational models \cite {Nowak98} and laboratory studies \cite{Mili02} \cite{Piazza08}, emphasize the role of reputation in supporting the evolution and the maintenance of cooperation. Experimental studies mentioned confirm that humans are able to maintain high levels of cooperation through the indirect reciprocity mechanisms offered by it. Reputation indeed allows to identify the cooperators and, at the same time, to exclude non-cooperators through social control \cite{Oto04} \cite{Giardini12}. This information through communication can flow freely. 
Indeed we can obtain informations about the reliability (\textit{i.e.}, the reputation) of a partner even without previous interactions with that individual and adjust our behaviour consequently. Reputation and communication are therefore intimately related. The language and the exchange of socially relevant information appears essential to ensure the functioning of the human cooperation model \cite{Dunbar04}. We know that reputation can improve material prosocial behaviour (\textit{e.g.}, a fairer allocation in a dictator game),  however, is less clear the relationship between reputation and informative behaviours, which are fundamental for the transmission and propagation of the reputational information. Feinberg and colleagues \cite{Feinbergossip} have shown that individuals are ready to share information on non-cooperative individuals, even in situations where this action results expensive. Humans seem to have a strong tendency to transmit social evaluations, however the studies conducted so far have examined this behaviour only in contexts where there was no particularly reason to omit information. Indeed, participants were not competing with each other.
Today, more than ever before, information and communication technologies connect people around the globe allowing them to exchange information and to work together, overcoming the physical separation constraints. These new opportunities for large-scale interaction, as well as the chance to make accessible our opinion to the community of Internet users (\textit {i.e.}, bi-directionality), allowed the development of systems based on online feedback mechanisms. Actually we are witnessing the proliferation of services that rely on reputation systems \cite{Della03}. The exchange of information and the presence of "reputation systems" ensure the functioning of e-commerce sites, such as Amazon and e-Bay, as well as services like Tripadvisor.  Given the expansion of communication possibilities introduced by ICTs, it become necessary to understand how reputation and gossip affect cooperation and competition dynamics in virtual environment, as well as to verify how cooperative informative behaviours change within a competitive frame. 

\section{Overview of Present Studies}
It was settled a 2x2 design in which reputation and cost of gossip could be present or absent. Four conditions were identified: 

\begin{itemize}
\renewcommand{\labelitemi}{$\bullet$}
\item Condition 1 (Reputation system ON, Cost of gossip OFF)
\item Condition 2 (Reputation system OFF, Cost of gossip OFF)
\item Condition 3 (Reputation system ON, Cost of gossip ON)
\item Condition 4 (Reputation system OFF, Cost of gossip ON)

\end{itemize} 

According to this, three studies were conducted. The first saw the activation of the conditions 1 and 2, the second one considered the condition 3 and 4, and the third compared the conditions in which the cost of gossip was present (\textit {i.e.}, Condition 3 and 4) and those where it was absent (\textit {i.e.}, Condition 1 and 2). 

\section{Participants}

The first study (condition 1 and 2) involved 72 volunteers (38 females), with an average age of 22 (s.d. 3, 7). 
Instead, a total of 174 participants (129 females), with an average age of 22 (s.d. 4,7), were engaged in the study 2 (condition 3 and 4).
A sample consisting of 246 individuals (167 females), derived from previous studies, was selected for the analyses concerning the study 3 (condition 1 and 2 vs. condition 3 and 4).
For each study the participants were recruited through complete voluntary census.

\section{Methods ad Procedures}

\textbf{Trustee Game}.
Taking inspiration from the most famous social dilemma games (e.g., Ultimatum Game, Trust Game) we have developed a multiplayer virtual game called Trustee Game. The game was realized through \textit{Google Apps}, using the \textit{Google Script} programming language. Within Trustee Game, groups formed by 6 players interacted anonymously with the instruction to win the game for a total time of 30 minutes and 45 turns. Each player has had three types of resources. One was equal to 50 units (i.e., maximum resource) and the other two constituted the minimal resources (i.e., 5 units for each). The maximum resource was always different for each player and it was assigned at random. Each participant covered all the roles in the game for the same number of times (i.e., 15) and leaded for each role 3 interactions with each other member within the group. Moreover, the interaction sequence was random. The player with the highest minimum resource after 45 rounds achieved the victory. \\

\textbf{Roles}. 

\textit{Donor}: has the task to make an offer and a request to the receiver. The donor offers his greatest resource, among the three at his disposal, and asks in return his minimum resource to the receiver. Using the sliders the donor chooses how much resource to offer and the amount to ask in return. Once established the terms of trade the donor should click on the "Go" button within 10 seconds.\\

\textit{Observer}: has to judge the donor's action. The observer has a clear vision of the exchange proposed. Indeed, the observer is able to display both the amount and the type of resources involved in the deal. In addition, the observer can provide a hint to the receiver, clicking on the button "suggest to accept", "no hint" or "suggest to refuse". The time available for the observer to make his choice is 10 seconds.\\

\textit{Receiver}: can only see the amount and the type of the resource offered by donor. Indeed, he is unaware of what the donor asked in return. The receiver may decide to "accept" or "reject" the donor deal without other information, or may require the observer suggestion (by clicking on "ask suggestion" button). When reputation system is active, the receiver can also see the rating (i.e., the number of like and dislike accumulated) of the observer with which he interacts. After his decision about the exchange offer, the receiver becomes completely aware of the donor request. If the receiver asked the suggestion and the like system is active, he has the opportunity to reward or punish (i.e., give a like or a dislike) the observer. The receiver has 18 seconds to make his own decisions. \\

\textbf{Setting.} 
The experiments took place within the computer lab of the School of Psychology of Florence. Each computer station was isolated from the others through separators in order to avoid interactions outside the game. Once arrived, participants were instructed about the rules of the game and were invited to clear up their doubts about the game mechanics. After that the game was launched on the machines.

\subsection*{Data Analysis}
The preconditions necessary to inferential analyzes were verified on the data produced by the experiments. For all the continuous variables that were under investigation, the normality of the distribution was assessed through the analysis of asymmetry and kurtosis values. Also, were verified the presence of an adequate sample size in order to obtain robust statistics. On continuous variables that do not respect the preconditions a discretization were made, using the median as a reference, and thus defining two levels for each variable. The analyzes on these parameters were conducted using the Pearson's chi-square test.

\section{Results}

\begin{table}[]
\centering
\caption{Reputation effect on material prosocial behaviour}
\label{repeffect}
\begin{tabular}{llll}
\textit{Variables}                    & \multicolumn{1}{c}{\textbf{Reputation Off}} & \multicolumn{1}{c}{\textbf{Reputation On}} & \multicolumn{1}{c}{χ²}      \\ \hline
(Study 1) Difference Donation-Request & \multicolumn{1}{c}{34.5\% (+)}               & \multicolumn{1}{c}{43.2\% (+)}              & \multicolumn{1}{c}{29.25**} \\ \hline
(Study 2) Difference Donation-Request & \multicolumn{1}{c}{46.6\% (+)}              & \multicolumn{1}{c}{48.2\% (+)}             & \multicolumn{1}{c}{ns}      \\ \hline
\\ $^{**}p<.01$                                    
\end{tabular}
\end{table}

Within the environment characterized by a free information transmission, reputation seems to induce donors to propose fairer deals (i.e., more positive difference between amount offered and asked). However, when the communication involves a cost, reputation fails to influence a prosocial allocation behavior (Table \ref{repeffect}). 

\begin{table}[]
\centering
\caption{Reputation effect on informative prosocial behaviour}
\label{repeffect2}
\begin{tabular}{llll}
\textit{Variables}             & \multicolumn{1}{c}{\textbf{Reputation Off}} & \multicolumn{1}{c}{\textbf{Reputation On}} & \multicolumn{1}{c}{$\chi^2$}      \\ \hline
(Study 1) Suggestion Coherence & \multicolumn{1}{c}{54.9\% (+)}              & \multicolumn{1}{c}{54.4\% (+)}             & \multicolumn{1}{c}{ns}      \\ \hline
(Study 2) Suggestion Coherence & \multicolumn{1}{c}{29.7\% (+)}              & \multicolumn{1}{c}{39.1\% (+)}             & \multicolumn{1}{c}{35.66**} \\ \hline
 \\ $^{**}p<.01$  
\end{tabular}
\end{table}

The presence of reputation mechanisms does not seem to influence the goodness of the suggestion provided in first study (i.e., where information flow freely). Whereas, the reputational influence on informative conducts occurs in the second study (i.e., where the suggestion involves a cost)(Table \ref{repeffect2}). 
  
\begin{table}[]
\centering
\caption{Capability of reputation to identify those who behave cooperatively}
\label{repcap}
\begin{tabular}{lccc}
\textit{Variables}             & \textbf{Bad Reputation} & \textbf{Good Reputation} & $\chi^2$ \\ \hline
(Study 1) Suggestion Coherence & 47.6\% (+)              & 55.3\% (+)               & 9.60**               \\ \hline
(Study 2) Suggestion Coherence & 27.5\% (+)              & 49.8\% (+)               & 77.09**              \\ \hline
(Study 2) Suggestion provided  & 55.7\% (0)              & 33.7\% (0)               & 71.86**              \\ \hline    
 \\ $^{**}p<.01$                           
\end{tabular}
\end{table}

In both studies, reputational mechanisms show themselves able to identify who acts in a prosocial manner as observer. Indeed, those who provide coherent suggestions more frequently earn a good reputation. This effect is more pronounced in the study 2, where the communication cost drastically reduces the number of suggestions in the system. This reduction of information occurs both in situations with ($\chi^{2}=107.05, p<.01$) and without ($\chi^{2}=225.28, p<.01$) reputation mechanisms . Reputation also identifies those who provide an expensive suggestion to the receiver. Indeed, individuals with good reputation abstain themselves less frequently from providing a costly suggestion than those who have obtained a bad one (Table \ref{repcap}).

\begin{table}[]
\centering
\caption{Prosocial informative behaviour response time}
\label{responsetime}
\begin{tabular}{lccc}
\textit{Variables}                & \textbf{Bad Reputation}                                                            & \textbf{Good Reputation}                                                       & $\chi^2$                           \\ \hline
(Study 1) Time to take a decision & 11.7\% (+)                                                                         & 8.6\% (+)                                                                      & 3.98*                        \\ \hline
(Study 2) Time to take a decision & 8.5\% (+)                                                                          & 5.7\% (+)                                                                      & 3.83*                        \\ \hline
                                  &                                                                                    &                                                                                &                              \\
                                  & \textbf{\begin{tabular}[c]{@{}c@{}}Costly suggestion \\ not provided\end{tabular}} & \textbf{\begin{tabular}[c]{@{}c@{}}Costly suggestion \\ provided\end{tabular}} & \multicolumn{1}{l}{}         \\ \hline
(Study 3) Time to take a decision & 13.4\% (+)                                                                         & 1.4\% (+)                                                                      & \multicolumn{1}{l}{131.32**} \\ \hline
 \\ $^{*}p<.05$; $^{**}p<.01$ 
\end{tabular}
\end{table}
Both in Study 1 and in Study 2 those with a good reputation (i.e., those who mainly act in a prosocial manner in the observer role), use less time to take a decision. Within the Study 3, the same trend (i.e., minor decision time) is recorded among participants who provide an expensive suggestion and those who does not perform such action. As we can see from the table \ref{responsetime}, those who decide to pay a cost in order to transmit an evaluation to the receiver, use a significantly shorter time.  

\begin{table}[]
\centering
\caption{Relationship between reputation and feedback behaviour in condition 1}
\label{inertia1}
\begin{tabular}{ccccc}
\hline
\multicolumn{1}{l}{}                                                               & \multirow{2}{*}{\begin{tabular}[c]{@{}c@{}}Coherence\\ on like\end{tabular}} & \multicolumn{2}{c}{Reputation} & $\chi^2$                       \\ \cline{3-5} 
\multicolumn{1}{l}{}                                                               &                                                                              & Good           & Bad           &                          \\ \cline{1-4}
\multirow{3}{*}{\begin{tabular}[c]{@{}c@{}}Suggestion \\ Coherent\end{tabular}}    & Coherent                                                                     & 56             & 10            & \multirow{3}{*}{44.11**} \\ \cline{2-4}
                                                                                   & No like                                                                      & 34             & 13            &                          \\ \cline{2-4}
                                                                                   & Not Coherent                                                                 & 6              & 23            &                          \\ \hline
\multirow{3}{*}{\begin{tabular}[c]{@{}c@{}}Suggestion\\ Not Coherent\end{tabular}} & Coherent                                                                     & 8              & 11            & \multirow{3}{*}{10.40*}  \\ \cline{2-4}
                                                                                   & No like                                                                      & 19             & 7             &                          \\ \cline{2-4}
                                                                                   & Not Coherent                                                                 & 17             & 4             &                          \\ \hline
 \\ $^{*}p<.05$; $^{**}p<.01$ 
\end{tabular}
\end{table}

\begin{table}[]
\centering
\caption{Relationship between reputation and feedback behaviour in condition 3}
\label{inertia2}
\begin{tabular}{ccccc}
\hline
\multicolumn{1}{l}{}                                                               & \multirow{2}{*}{\begin{tabular}[c]{@{}c@{}}Coherence\\ on like\end{tabular}} & \multicolumn{2}{c}{Reputation} & $\chi^2$                       \\ \cline{3-5} 
\multicolumn{1}{l}{}                                                               &                                                                              & Good           & Bad           &                          \\ \cline{1-4}
\multirow{3}{*}{\begin{tabular}[c]{@{}c@{}}Suggestion \\ Coherent\end{tabular}}    & Coherent                                                                     & 131            & 28            & \multirow{3}{*}{27.09**} \\ \cline{2-4}
                                                                                   & No like                                                                      & 183            & 130           &                          \\ \cline{2-4}
                                                                                   & Not Coherent                                                                 & 26             & 14            &                          \\ \hline
\multirow{3}{*}{\begin{tabular}[c]{@{}c@{}}Suggestion\\ Not Coherent\end{tabular}} & Coherent                                                                     & 21             & 37            & \multirow{3}{*}{13.04**} \\ \cline{2-4}
                                                                                   & No like                                                                      & 73             & 63            &                          \\ \cline{2-4}
                                                                                   & Not Coherent                                                                 & 19             & 5             &                          \\ \hline
 \\ $^{**}p<.01$ 
\end{tabular}
\end{table}

Another interesting result emerges from the relationship between the like/dislike action (i.e., the social feedback) of the receiver and the reputation of the observer. In all the conditions in which the reputation mechanisms were active, an observer with a good reputation (i.e., more likes than dislikes) attracted more frequently a reward (i.e., likes) apart from the fact that the suggestion provided was good or bad. Specularly, an observer with a bad reputation received more frequently a dislike, even when he provided a coherent suggestion (Table \ref{inertia1}, Table \ref{inertia2}).

\section{Discussion}
The experiments showed that in a virtual social dilemma situation, reputation could trigger cooperative behaviours. The type of prosocial conduct that is elicited by reputation (i.e., material or informative) depends on the amount of available information. When the suggestion does not involve any penalty, and therefore there is a greater possibility that the observer provides an assessment of the deal, the reputation seems to be able to influence the donor behaviour. Instead, when the amount of information decreases due to the introduction of a cost to provide a suggestion, donors no longer appear to be influenced by reputation. However, this cost seems to act as a filter for those that do not provide coherent suggestions. Indeed, in an environment characterized by an expensive transmission of information, reputation appears to push the observers to provide more coherent advice. \\
Another result highlighted by this work concerns the reputation ability to efficaciously identify those who behave in a prosocial manner. Indeed, those who earn a good reputation act in a more prosocial way (i.e., provide good suggestion, pay a cost to trasmit information). The fact that those who perform prosocial actions employ a smaller amount of time, seems to confirm Rand and colleagues results \cite{Rand14}. Indeed, cooperative decision making seems to rely on social heuristics (i.e., fast and prosocial decision rules) which are made salient by reputation.  \\
However, reputation also influences people's decision making in a non-trivial manner. As we have seen, reputation once acquired tends to perpetuate itself apart from the goodness of the suggestions. Through gossip the social prototype (good partner or bad one) is transmitted but at the same time, this prototype ends up influencing the decision making of individuals in the direction of its confirmation. Reputational systems are frequently used in virtual environment to ensure the functionality of online services. Nevertheless, their use is often unaware of the limits with which reputation can be applied. The future perspective is therefore to create adaptive automatic systems that will correct the distortion made by reputation, allowing a true estimate of people online behaviours as well as to be aware of the real value of a good or of a company. Indeed, the information conveyed by the reputation could not be connected to the real behaviour and so induce people to make mistakes (e.g., trust the wrong social partner). In addition, through the study of reputation dynamics the present work aims to contribute to the efficiency and effectiveness of virtual platforms, by means of an accurate modeling of their psychosocial ergonomy. Finally, allowing to design virtual envrionments able to elicit prosocial (e.g., energy consumption reduction) and health (e.g., support for addictions) behaviours from users.  

A limit of the present work lies in the low power (i.e., too few participants), and in the high homogeneity of the results. Therefore, future research should deal with this aspect. 

\section{Acknowledgement}
This work was supported by EU Commission (FP7-ICT-2013-10) Proposal No. 611299 SciCafe 2.0.

\end{document}